\documentclass[structabstract]{aa}

\usepackage{graphicx}
\usepackage{txfonts}
\usepackage[authoryear]{natbib}
\bibpunct{(}{)}{;}{a}{}{,}
\def\simless{\mathbin{\lower 3pt\hbox
     {$\rlap{\raise 5pt\hbox{$\char'074$}}\mathchar"7218$}}}   
\def\simmore{\mathbin{\lower 3pt\hbox
     {$\rlap{\raise 5pt\hbox{$\char'076$}}\mathchar"7218$}}}   

\def\msun{~{\rm M}_\odot}

\begin{document}
\title{
The time-lag -- photon-index correlation in GX 339--4
}

\subtitle{}

\author{
Nikolaos D. Kylafis\inst{1,2} 
\and
Pablo Reig\inst{2,1}
}

\institute{
University of Crete, Physics Department \& Institute of
Theoretical \& Computational Physics, 70013 Heraklion, Crete, Greece\\
\and
IESL \& Institute of Astrophysics, Foundation for Research and Technology-Hellas, 71110 Heraklion, Crete, Greece
}

\date {Received ; Accepted ;}


\abstract
{
Black-hole transients, as a class, exhibit during their outbursts
a correlation between the time lag 
of hard photons with respect to softer ones and the photon index 
of the hard X-ray power law. The correlation is not very tight and 
therefore it is necessary to examine it source by source.}
{The objective of the present work is to investigate in detail the 
time-lag -- photon-index
correlation in GX 339-4, which is the best studied black-hole transient.
}
{
We have obtained {{\it RXTE}} energy spectra and light curves 
and have computed  the photon
index and the time lag of the $9 - 15$ keV photons with respect  to the $2 -
6$ keV ones.  The observations cover the  first stages of the hard state,
the pure hard state,  and the
hard-intermediate state.
}
{
We have found a tight correlation between time lag and photon index $\Gamma$ in the 
hard and hard-intermediate states.  At low $\Gamma$, the 
correlation is positive and it becomes negative at large $\Gamma$.  By 
assuming that the hard X-ray power law index $\Gamma$ is produced by 
inverse Compton scattering of soft disk photons in the jet, we have 
reproduced the entire correlation by varying two parameters in the jet:  
the radius of the jet at its base $R_0$ and the Thomson optical depth 
along the jet $\tau_\parallel$.  We have found that, as the luminosity 
of the source increases, $R_0$ initially increases and then decreases. 
This behavior is expected in the context of the Cosmic Battery.
}
{
Our jet model nicely explains the correlation with reasonable values of the 
parameters $R_0$ and $\tau_\parallel$. These parameters also correlate between 
themselves.  As a further test of our model, we predict the break frequency in 
the radio spectrum as a function of the photon index during the rising part 
of an outburst.
}


\authorrunning{Kylafis \& Reig 2018}

\titlerunning{Timelag - spectral-index correlation in GX 339-4}

\maketitle


\section{Introduction}

Despite the fact that significant progress has been made in the phenomenology
and modelling of the X-ray emission of black-hole transients (BHTs)  in the last
decade,  the accretion/ejection phenomena in these systems  are still poorly
understood. Understanding the emitted spectra is extremely important, because it
reveals the physical processes giving rise to the high-energy radiation.
However, at present, the knowledge that we can obtain is limited because there
are more ways than one to reproduce the observed spectra
\citep{titarchuk94,esin97,poutanen99,reig03,markoff05,done07}.  Timing
properties add a new view of the sources and may help breaking the degeneracy
inherent in the spectral analysis, but again there are various ways to
explain the observations
\citep{poutanen99,bottcher99,kotov01,reig03,arevalo06,kroon16}.  

The way forward must come from the combination of spectral and timing
information.  Stringent constraints to the models come from the correlated
timing and spectral behavior
\citep{pottschmidt03,shaposhnikov09,reig13,stiele13,shidatsu14,grinberg14,kalamkar15,
altamirano15,reig15,reig18,axelsson18} Hence, to make progress in our
understanding of the accretion/ejection phenomena in BHTs, the observations 
must uncover these correlations  and the models must explain them and make
predictions that can either verify or disprove the models.

Recently, we examined the BHTs as a class and found a correlation between  the
photon number spectral index $\Gamma$ of the hard X-rays and the timelag $t_{\rm
lag}$ of the 9 - 15 keV photons with respect to the  2 - 6 keV ones
\citep{reig18}.  Although the correlation is statistically significant, it shows
large scatter, because the geometry and the physical properties of the
comptonizing region are not the same in all BHTs. Consequently, a closer
examination of each individual source may give more insights into
the ongoing processes. We will address the scatter of the correlation in a
subsequent paper (Reig \& Kylafis, in preparation). In this work we start the
individual investigation of the sources with GX 339-4. This source has exhibited
several outbursts and it is one of the best studied BHTs.

The structure of this Letter is as follows: in \S~2 we present the observations
and the data analysis,  in \S~3 we describe briefly our jet model, in \S~4 we
compare the model with the observations and make a  theoretical prediction, in
\S~5 we comment on our findings, and in \S~6 we draw our conclusions.

\section{Observations and data analysis}

\begin{figure}
\centering
\begin{tabular}{c}
\includegraphics[angle=0,width=7cm]{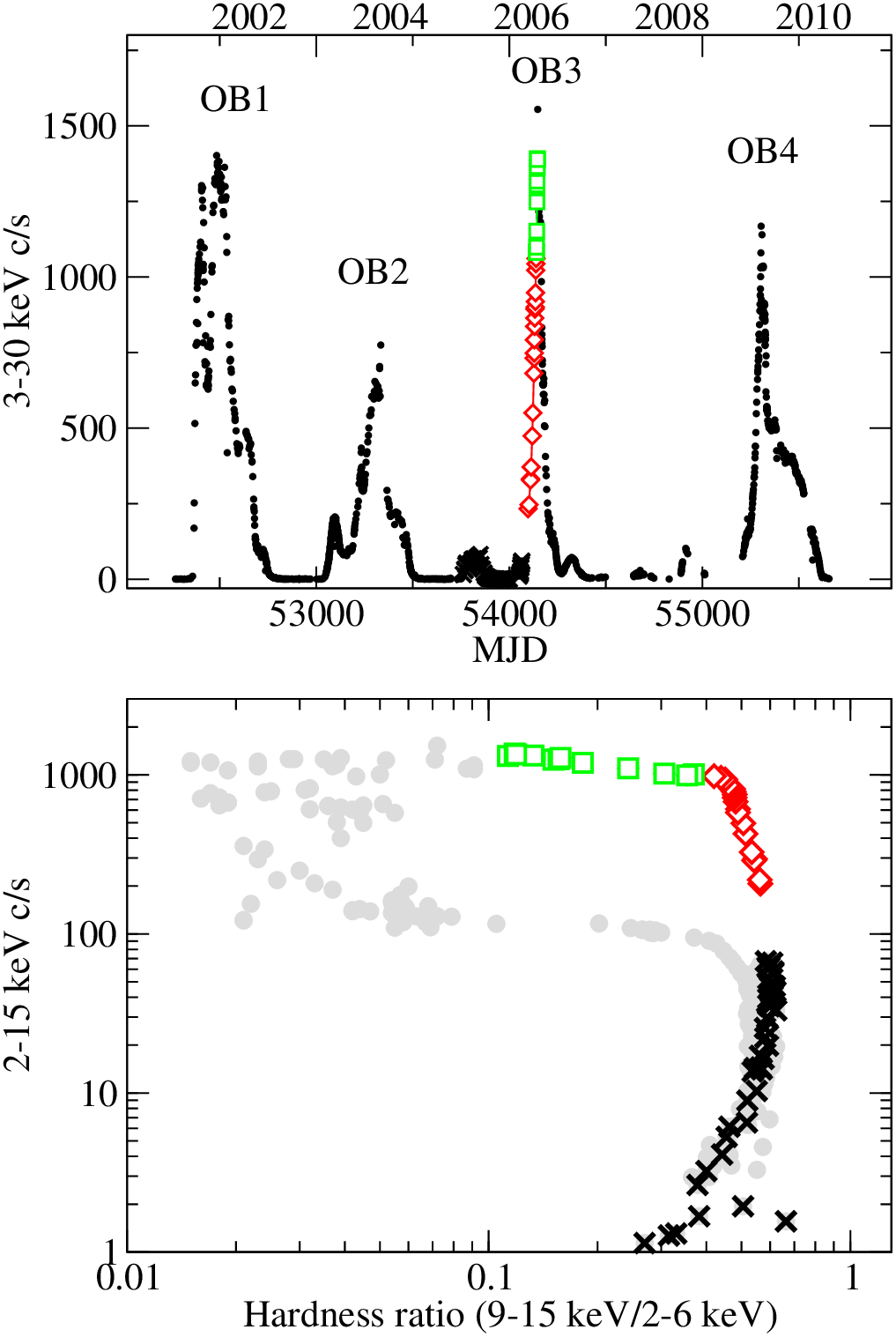} \\
\includegraphics[angle=0,width=7cm]{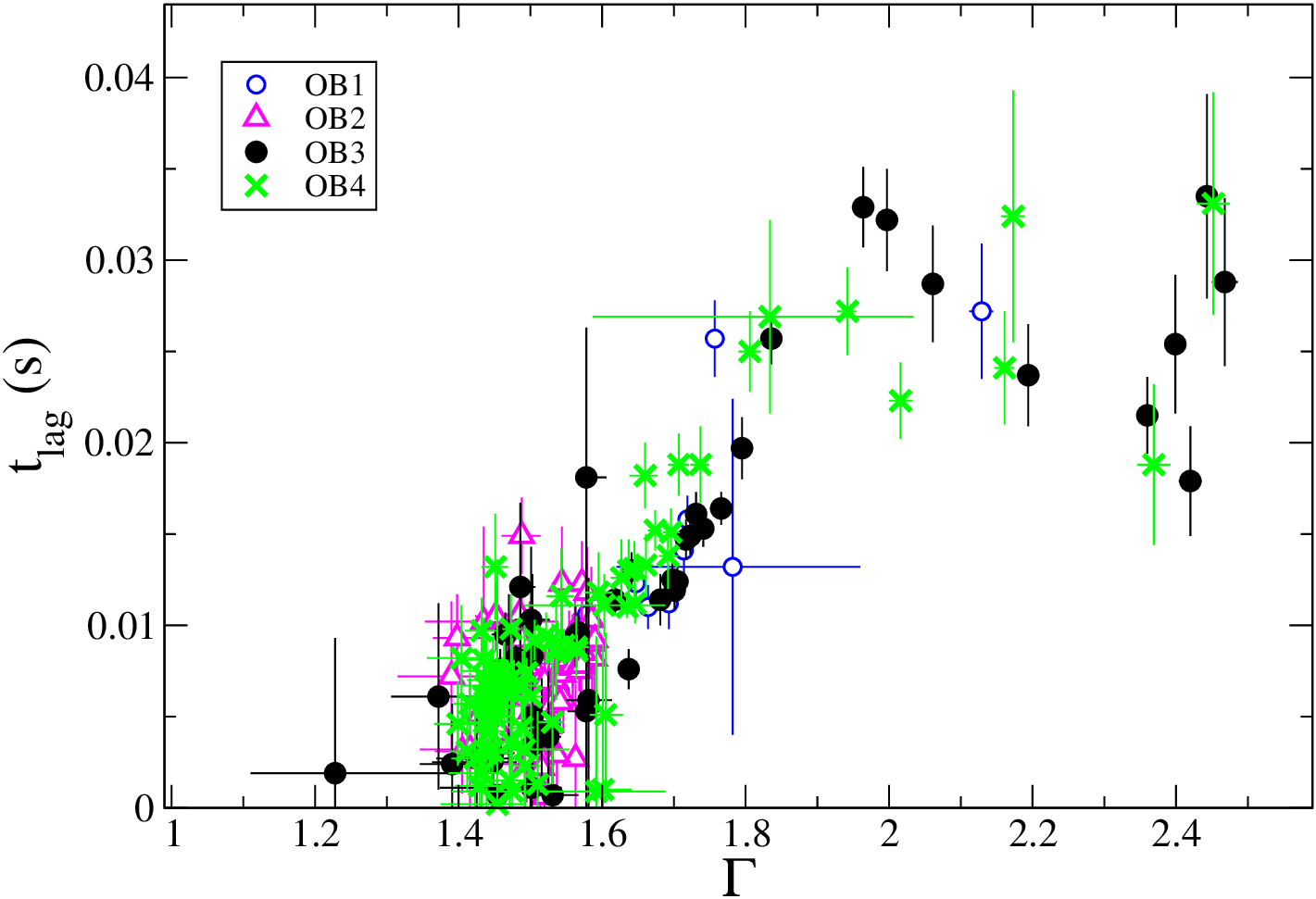} \\
\end{tabular}
\caption{{\em Top panel}: Long-term light curve of GX 339--4 showing four major
X-ray oubutbursts. {\em Middle panel}: Hardness-intensity diagram for
the 2006-2007 outburst. 
{\em Bottom panel}: Time lag between $9-15$ keV photons with
respect to $2-6$ keV photons against the photon index of the hard
power-law continuum. All four outbursts are included.}
\label{fig1}
\end{figure}

This study was performed using archival data of GX 339--4 from the Rossi X-ray
Timing Experiment ({\it RXTE}). During the time that the mission was
operational, GX 339-4 exhibited four major X-ray outbursts (Fig.~\ref{fig1}, top
panel). In this Letter, we focus on the 2006--2007 outburst because it presents
the best statistics and the best sampled hard-intermediate state, which allow us
to explore the upper branch of the $q$-diagram in detail (Fig.~\ref{fig1},
middle panel). However it should be noted that the shape of the correlation 
examined in this Letter is
similar in all outbursts, as it can be seen in the bottom panel of
Fig.~\ref{fig1}  \citep[see also Fig. 6 in][]{altamirano15}.

The details of the data analysis can be found in \citet{reig18}. Here we briefly
summarize the data products that we have used. To compute the time lags, we
generated light curves $x_i(t)$ with time bin size $2^{-7}$ s in the energy
ranges 2--6 keV (soft band) and 9--15 keV (hard band). We calculated an average
cross  vector $<C(\nu_j)>=<X_1^*(\nu_j)X_2(\nu_j)>$, where $X_i(\nu_j)$ are the
Fourier transforms of light curves $x_i(t)$ and the asterisk denotes complex
conjugate. The average was perfomed over multiple adjacent 64-s  segments. The
final time lag of the $9-15$ keV photons with respect to the $2-6$ keV ones
resulted from the average of the time lags in the frequency range 0.05--5 Hz.
The selection of the energy and frequency ranges was driven by the
instrument sensitivity and signal-to-noise.  

The spectral analysis was performed in the energy range 2--25 keV\footnote{
In \citet{reig18}, we used both PCA and HEXTE data. The power-law index there is
systematically lower by $\sim$0.2 due to the different energy range used in the
fits.}. We  obtained
the energy spectra using the standard-2 mode of the  {\it RXTE} PCA instrument.
To characterize the spectral continuum, we used an absorbed broken power law
model. The hydrogen column density was fixed to $N_H=4 \times 10^{21}$ cm$^{-2}$
\citep{dunn10}. A narrow  Gaussian component (line width $\sigma \simless 0.9$
keV) was added to account for the iron emission line at around 6.4 keV. The
photon index used in our analysis is the one that corresponds to the hard power
law, that is, after the break.

\section{Jet model}

Over the years, we have developed a simple jet model that explains  
the spectral and timing observations of BHTs quite well
\citep{reig03,giannios04,giannios05,kylafis08,reig15,reig18}. The model used
here is the same as that of \citet{reig15} and \citet{reig18}.  The reader is
referred to these papers for details. Here, we give only a brief description.

We assume a parabolic jet with a finite acceleration region.  The  electrons in
the jet have a component of their velocity $v_{\parallel}$ parallel to the axis
of the jet and one perpendiculat $v_{\perp}$ to it.  This is justified if the
electrons in the jet have a steep power-law distribution in their Lorentz
factors \citep{giannios05}.

The accretion flow around the black hole  consists of an inner hot flow and an
outer  geometrically thin accretion disk \citep{esin97,kylafis15b}.   The jet is
fed by the hot inner  flow, while the thin accretion disk is the source of
blackbody photons at the base of the jet. These soft photons, either escape
uscattered or are scattered in the jet and have on average their energy
increased. This upscattering of the soft blackbody photons produces the hard
X-ray power law with index $\Gamma$.  Furthermore, this upscattering causes an
average timelag of the harder photons with respect to the softer ones.

The parameters of the model are: the optical depth $\tau_{\parallel}$ along
the axis of the jet, the radius of the jet $R_0$ at its base,  the minimum
Lorenzt factor\footnote{The electrons in jets are believed to obey a
power-law distribution in Lorentz $\gamma$  ($dN/d\gamma \propto
\gamma^{-\alpha}$), from some $\gamma_{\rm min}$ to some $\gamma_{\rm max}$. 
Because of the steepness of this distribution ($\alpha \simmore 2$), most of the
electrons in the jet have Lorentz gamma factors close to $\gamma_{\rm min}$. 
For simplicity, we have taken all the electrons to have $\gamma = \gamma_{\rm
min}$. } $\gamma=1/\sqrt{1-(v_0^2+v_{\perp}^2)/c^2}$ of the electrons in the
jet, the distance $z_0$ of the bottom of the jet from the black hole, the total
height $H$ of the jet, the temperature  $T_{bb}$ of the soft-photon input and 
the size $z_1$ and exponent $p$ of the acceleration zone  $v_{\parallel}(z)
=(z/z_1)^p ~ v_0$, for $z \le z_1$,  and $v_0=v_{\parallel}(z)$ for
$z>z_1$. The energy spectra, from which we derive the photon index $\Gamma$,
and the time lag $t_{\rm lag}$ have been computed  by Monte Carlo
\citep{reig15,reig18}.

\begin{figure}
\centering
\includegraphics[angle=0,width=8cm]{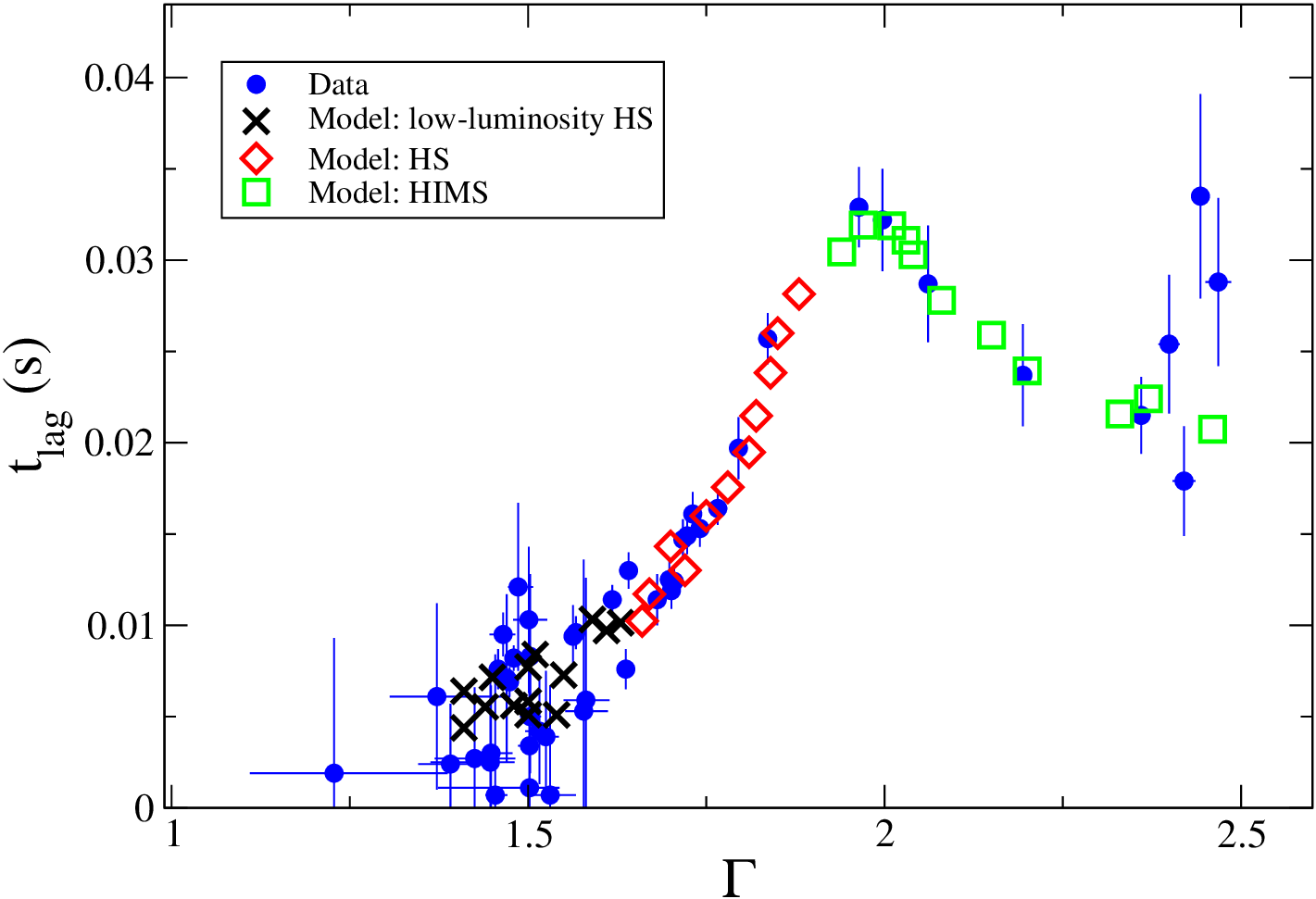}
\caption{Time lag of the $9 - 15$ keV photons with respect to the $2 - 6$ keV 
ones as a function of the spectral index $\Gamma$ of the hard X-rays
for the 2006-2007 outburst.  The blue circles give the 
observations and the different symbols give the model results, as follows:
black crosses - low-luminosity HS, red diamonds - HS, and
green squares - HIMS.}
\label{fig2}
\end{figure}

\section{Comparison of the model with the observations}

In Fig.~\ref{fig2} (see also the bottom panel of Fig.~\ref{fig1}),  we show the
average time lag $t_{\rm lag}$ as a function of the spectral index $\Gamma$ for
the 2006-2007 outburst.  The filled circles give the observations and the rest 
of the symbols represent the model results as follows: For $\Gamma \simless 
1.6$ the source is considered to be in the  first stages of the hard
state  (black crosses in Fig.~\ref{fig2}),  for $1.6 \simless \Gamma \simless
1.9$ the source is taken to be in the hard state (HS, red diamonds in
Fig.~\ref{fig2}),  and for $\Gamma \simmore 1.9$ the source is in the
hard-intermediate state (HIMS, green squares in Fig.~\ref{fig2}). Here, we
follow the classification of \citet{belloni10} of the spectral states of BHTs.
To facilitate a comparison with the position of the source in the
hardness-intensity diagram, we use the same symbols as in the middle panel of
Fig.~\ref{fig1}. We note that while the symbols represent the results of our
model in Fig.~\ref{fig2}, they correspond to the actual observations in
Fig.~\ref{fig1} (top and middle panels). 

To reproduce the observations, we have varied only two parameters, the
radius $R_0$  of the jet at its base and the Thomson optical depth 
$\tau_{\parallel}$ along the axis  of the jet.  The rest of the parameters are 
kept at their reference values \citep{reig18}, namely $z_0 = 5 r_g$, $H=10^5
r_g$, $v_0=0.8 c$, $\gamma=2.24$, $z_1= 50 r_g$, $p=0.5$, and $T_{bb} = 0.2$
keV. $r_g=GM/c^2$ is the gravitational radius. We assume a black-hole mass of 10
$\msun$.

\begin{figure}
\centering
\includegraphics[angle=0,width=8cm]{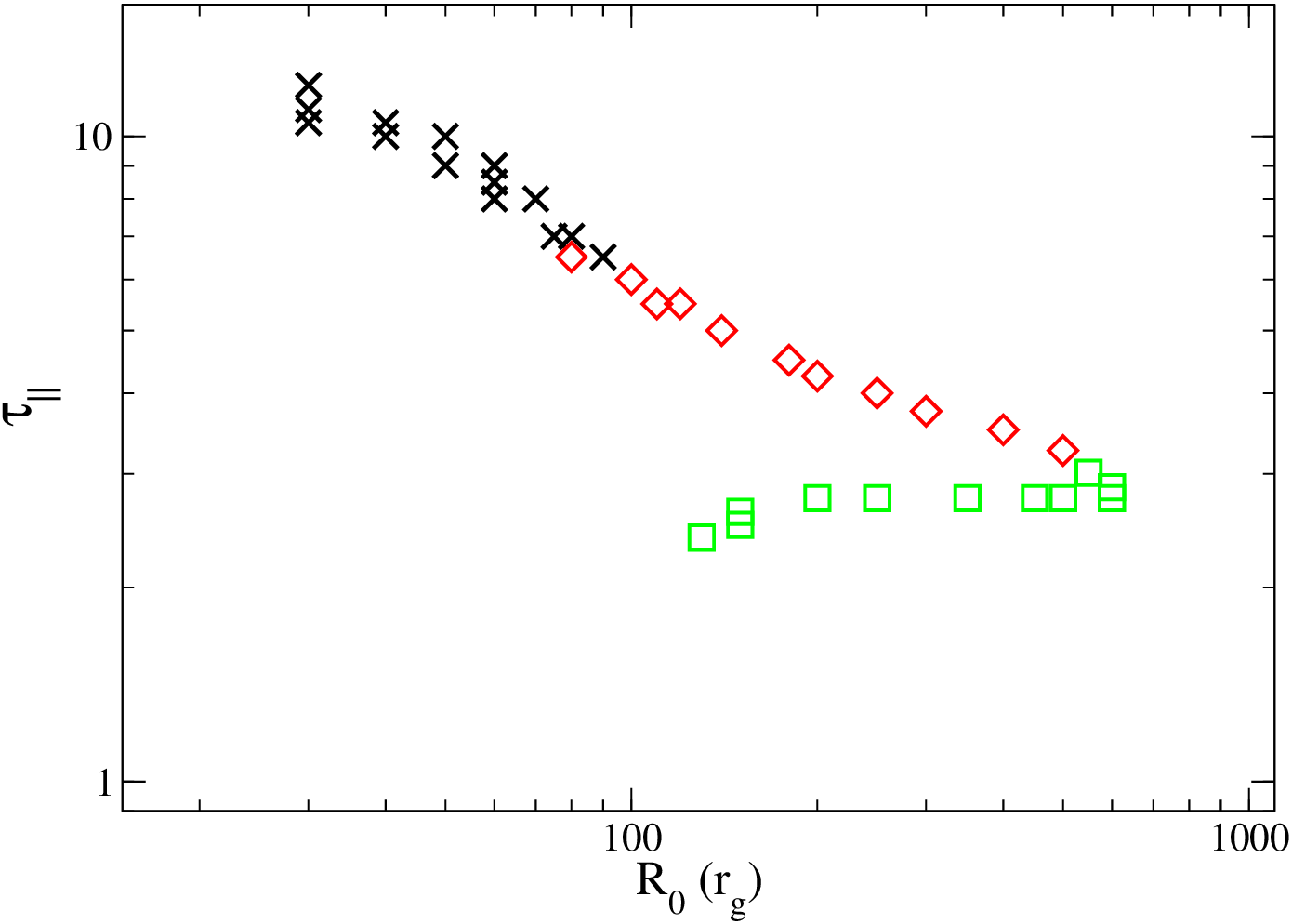}
\caption{Values of the parameters $R_0$ and $\tau_{\parallel}$
that correspond to the same models used in Fig.~\ref{fig2}.
The symbols are the same as in Fig.~\ref{fig2}.} 
\label{fig3}
\end{figure}

In Fig.~\ref{fig3}, we show the values of the model parameters $R_0$ and
$\tau_{\parallel}$, that we have used in Fig.~\ref{fig2}.  We use the same
symbols as in Fig.~\ref{fig2} for the three spectral states.  We see several
interesting results.  a) For the transitions from the low-luminosity HS to
the HS and  HIMS, the Thomson optical depth $\tau_{\parallel}$  decreases
monotonically from ~11 to ~ 2.25. b) For the same transitions, the radius $R_0$
of the base of the jet is not changing monotonically.  From the initial
stages of the outburst to the beginning of the HIMS $R_0$ increases, while for
the rest of the HIMS $R_0$ decreases. We will offer a physical explanation for
this behavior in \S~5. c) More importantly, the parameters $R_0$ and
$\tau_{\parallel}$ are correlated.   Given the simplicity of our model, the
match between observations and model is excellent and the values of the
parameters $R_0$ and $\tau_{\parallel}$ are quite reasonable.

\begin{figure}
\centering
\includegraphics[angle=0,width=8cm]{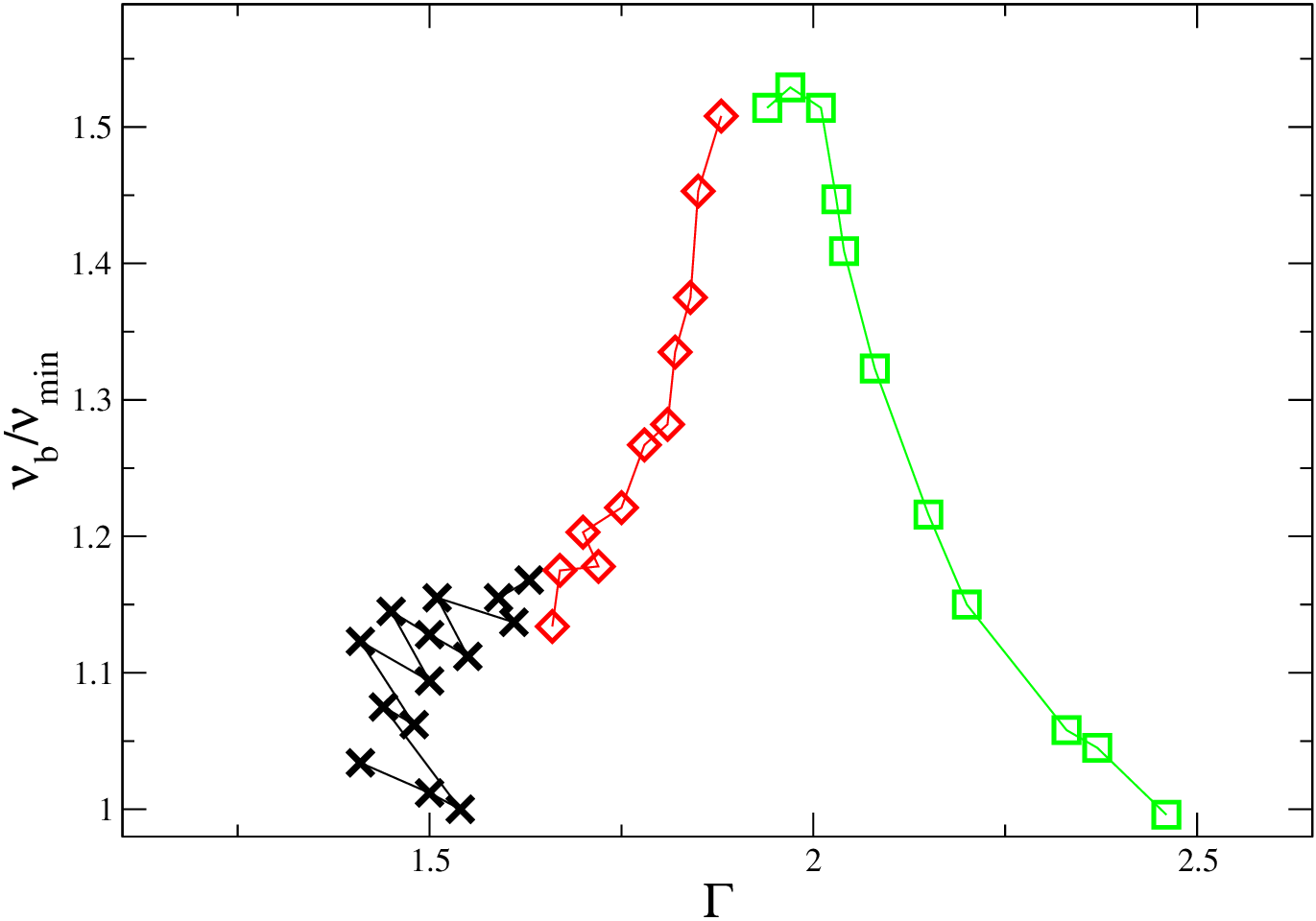}
\caption{Radio break frequency as a function of the photon index. The symbols
are the same as in Fig.~\ref{fig1}.} 
\label{fig4}
\end{figure}

Another success of our model is the fact that it can reproduce
qualitatively  the  correlation between the photon index of the power-law
component and the break frequency $\nu_{br}$ of the radio spectrum, i.e., the
frequency at which the entire jet becomes optically thin to synchrotron
radiation \citep{koljonen15}. In BHTs, this break is expected in the IR region
and could be related to the start of the particle acceleration in the jet
\citep{polko10}.  Fig.~\ref{fig4} shows $\nu_{br}$ as a function of $\Gamma$.
The break frequency was calculated following \citet{tsouros17} using the density
of the electrons at the base of the jet and the radius of the jet there (for the
jet models reported in Fig.~\ref{fig2}) and assuming that the magnetic field at
the base of the jet is $\sim 10^6$ Gauss \citep{giannios05}. The scaling for the
frequency is the minimum break frequency $\nu_{\rm min}$ obtained in the
low-luminosity HS. The symbols are the same as in Fig.~\ref{fig2}. The break
frequency in the rise of the  outburst should first increase and then decrease. 
The right branch of Fig.~\ref{fig4} agrees very well with the observations
reported in \citet{koljonen15} who found that the photon index anti-correlates
with the break frequency when the source transits from the HS to the HIMS (see
Fig. 2 in that reference). Unfortunately, there are not good enough data to
confirm the left branch, i.e. the dependence of the photon index on break
frequency when the source moves along the hard state. Our model predicts a
positive correlation between $\nu_{br}$ and $\Gamma$.


\section{Discussion}

The monotonic decrease of the optical depth $\tau_{\parallel}$  as $\Gamma$
increases is naturally expected in all Comptonization models.  Smaller optical
depth means smaller number of scatterings and therefore a steeper power law. On
the other hand, the initial increase and the later decrease of $R_0$ is more
challenging to explain.  Clearly, $R_0$ depends  crucially on the poloidal
magnetic field that is needed to eject the jet.

It is not easy to say what $R_0$ does as the luminosity of the source increases,
if the magnetic field is advected inwards from far away and it  is amplified on
its way by random processes, such as turbulence. Fortunately, it was recently
shown \citep{contopoulos18}
that the Cosmic Battery \citep{contopoulos98,contopoulos06} dominates
over all random processes in the creation of poloidal magnetic fields in
accreting black holes. Thus, we must use the Cosmic
Battery to explain the dependence of $R_0$ on luminosity.  There are no detailed
calculations on this yet, thus by necessity we will offer a qualitative
explanation.  A quantitative  one must be worked out and it will be done in the
near future.

Since the poloidal magnetic field produced by the Cosmic Battery  is
proportional to luminosity \citep{kylafis12},  at the beginning of the HS, 
with the luminosity low, the created poloidal magnetic field is  strong enough
to eject a jet only near the Inner Stable Circular Orbit  (ISCO). Thus, $R_0$ is
of order several gravitational radii. As the luminosity increases, the strength
of the poloidal magnetic  field increases and $R_0$ also increases, but not
indefinitely. This is because $R_0$ cannot increase beyond the transition radius
$R_{tr}$, between the hot inner flow and the outer thin accretion disk
\citep[see][for a recent determination of $R_{tr}$]{plant15}. Thus, somewhere at
the beginning of the HIMS, $R_0$ reaches $R_{tr}$ and after that it begins to
decrease because $R_{tr}$ decreases.

In Fig.~\ref{fig2}, time flows from left to right. While the duration of the
HS (red diamonds) is about 20 days, the time spent by the source in the HIMS 
(green squares) is
less than 4 days. According to our model, the width of the jet 
reduces by a factor of $\sim$3 in such a short time.
At the same time, the optical depth (i.e. density) 
along the axis of the jet remains about the same
(Fig.~\ref{fig3}). A dramatic shrinking of the jet, while its 
central density remains unaffected,
suggests that the jet quickly loses its outer
parts. This could happen if the Shakura-Sunyaev disk moves inward quickly
and ``scythes'' the jet.  In other words, the extension of the thin 
disk inwards shrinks the hot inner flow that feeds the jet and the outer parts
of the jet disappear, while its core remains more or less the same.
This idea seems to be in agreement with the reduction of the 
disk truncation radius in the HIMS \citep{plant15}.

\section{Conclusion}

In our recent work \citep{reig18}, we found that BHTs as a class  exhibit a
correlation between the power-law photon index  and the time lag between hard
and soft photons.  When considered as a whole (many sources), the correlation
shows a large amount of scatter, because different sources seem to obey somewhat
different correlations. Here, we have examined in detail this correlation for an
individual source, namely, GX 339-4. We have found that the correlation is very
tight. This correlation poses significant constraints on physical models of
BHTs. We have shown that we can explain the correlation using our simple jet
model and the Cosmic Battery. In addition, we make a prediction of how the
break frequency of the radio spectrum should vary during the rising part of a
large outburst.

\begin{acknowledgements}
We thank Ioannis Contopoulos for discussions regarding the Cosmic 
Battery and its effects on the size of the jet and Alexandros Tsouros for
computing the break frequencies.
We have also profited from useful discussions with Iossif Papadakis.
\end{acknowledgements}

\bibliographystyle{aa}
\bibliography{../bhb} 

\end{document}